\documentclass{ws-procs9x6}

\begin{document}

\title{THE SENSITIVITY OF\\
R-PROCESS NUCLEOSYNTHESIS\\
TO THE PROPERTIES OF NEUTRON-RICH NUCLEI}

\author{R. SURMAN$^*$, M.R. MUMPOWER, J. CASS, and A. APRAHAMIAN}

\address{Department of Physics, University of Notre Dame,\\
Notre Dame, IN 46556, U.S.A.\\
$^*$E-mail: rsurman@nd.edu\\
www.nd.edu}

\begin{abstract}
About half of the heavy elements in the Solar System were created by rapid neutron capture, or 
r-process, nucleosynthesis.  In the r-process, heavy elements are built up via a sequence of 
neutron captures and beta decays in which an intense neutron flux pushes material out towards 
the neutron drip line.  The nuclear network simulations used to test potential astrophysical 
scenarios for the r-process therefore require nuclear physics data (masses, beta decay 
lifetimes, neutron capture rates, fission probabilities) for thousands of nuclei far from 
stability.  Only a small fraction of this data has been experimentally measured.  Here we 
discuss recent sensitivity studies that aim to determine the nuclei whose properties are most 
crucial for r-process calculations.
\end{abstract}


\bodymatter

\section{Introduction}\label{sec:intro}

Current and upcoming radioactive beam facilities will have the capability to measure basic 
nuclear properties, such as masses and beta decay rates, of many neutron-rich nuclei for the 
first time.  These measurements have the potential to revolutionize our understanding of the 
evolution of nuclear structure far from stability. This understanding is crucial for many 
applications, including accurate models of the astrophysical processes in which these nuclei 
participate.

Rapid neutron capture, or r-process, nucleosynthesis is one such astrophysical process; see, e.g., 
Ref.~\refcite{arn07} for a recent review. In the r-process, heavy nuclei are built up by a sequence of fast 
neutron captures and beta decays, where the timescale for captures is initially much faster than that for 
decays\cite{cow91}.  Thus nuclei quite far on the neutron-rich side of stability participate in the 
process.  We observe the resulting r-process abundance pattern in the solar system and in many old stars, 
e.g., Ref.~\refcite{sne08}.  Exactly where astrophysically the r-process takes place, however, is still not 
conclusively determined\cite{arn07,qia07}.

Investigating potential astrophysical sites for the r-process involves simulations that require nuclear 
data for thousands of nuclei far from stability.  Since as of yet only a small percentage has been measured 
experimentally, we rely on theoretical models for this data. There is a long tradition of examining the 
impact of various choices of global nuclear models on shaping the r-process abundance pattern; see 
Refs.~\refcite{arn07,arc11} and references therein. Complementary to these efforts are r-process 
sensitivity studies, which aim to understand the role of {\em individual} pieces of nuclear data on 
r-process dynamics and to point out the nuclei whose properties play the most important role shaping the 
final abundance pattern.  The latter can potentially guide upcoming radioactive beam experiments by 
indicating the pieces of nuclear data with the greatest astrophysical impact.

Sensitivity studies have been performed for neutron capture rates\cite{beu08,sur09,sur11,mum12}, neutron 
separation energies\cite{bre12}, and beta decay rates\cite{cas12}.  Here we present a variation of the 
neutron separation energy study and examine the dependence of this and the beta decay studies upon the 
potential r-process astrophysical conditions.

\section{Sensitivity studies}\label{sec:sens}

We begin our sensitivity studies by choosing a baseline r-process simulation that produces a 
final abundance pattern that is a reasonable match to the solar pattern for $A>120$. We then 
modify one piece of nuclear data---a single mass or beta-decay rate, for example---and repeat 
the simulation.  The final mass fractions of this simulation $X(A)$ are compared to those of the 
baseline simulation $X_{\mathrm{baseline}}(A)$ using a global sensitivity measure $F$:
\begin{equation}
F=100\times \sum_A |X_{\mathrm{baseline}}(A)-X(A)|.
\label{eq:f}
\end{equation}
The mass fractions are related to the abundances $Y(A)$ by $X(A)=AY(A)$, and $\sum_{A} X(A)=1$. 
This process is repeated for every mass or beta decay rate, resulting in a sensitivity measure 
$F$ determined for each nucleus in the network.

For the r-process simulations we use the nuclear network code from Ref.~\refcite{sur01}. It is a 
dedicated r-process network code that includes neutron captures, photodissociations, beta decay, 
and beta-delayed neutron emission (and neutrino interactions and fission, though these options 
are not used for the sensitivity studies described here).  We use nuclear masses from 
Ref.~\refcite{mol95} (FRDM), Ref.~\refcite{duf95} (DZ), or Ref.~\refcite{gor10} (HFB-21), and 
neutron capture rates consistent with these mass sets from Refs.~\refcite{rau00}, or calculated with 
TALYS\cite{gor08}.  Beta decay rates are from Ref.~\refcite{mol03}.

\section{Binding energy study}\label{sec:be}

The study described here is based on the neutron separation energy sensitivity study of 
Ref.~\refcite{bre12}.  Neutron separation energies appear explicitly in the calculation of 
photodissociation rates via detailed balance:
\begin{equation} 
\lambda_\gamma(Z,A) \propto T^{3/2} \exp\left[-{\frac{S_n(Z,A)}{kT}}\right] \langle \sigma v \rangle_{(Z,A-1)}
\label{eq:gn}
\end{equation} 
where $T$ is the temperature, $\langle \sigma v \rangle_{(Z,A-1)}$ is the thermally-averaged 
neutron capture cross section for the nucleus with one less neutron, and $S_{n}(Z,A)$ is the 
neutron separation energy---the difference in binding between the nuclei $(Z,A)$ and $(Z,A-1)$.  
In Ref.~\refcite{bre12}, the neutron separation energies, as they appear in the equation above, 
were varied from their theoretical values by $\pm25$\%.  The sensitivity study proceeded as 
described in Sec.~\ref{sec:sens} above, for astrophysical conditions chosen to be similar to those of Ref.~\refcite{qia98}.

\begin{figure}[ht]
\begin{center}
\psfig{file=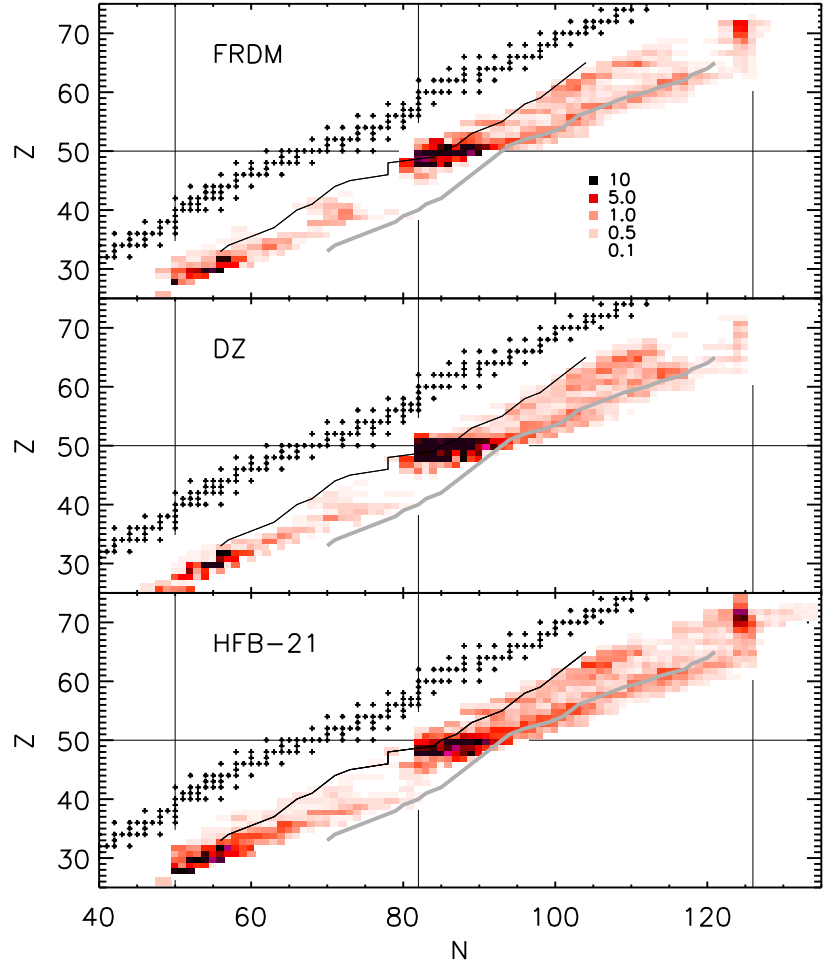,width=\textwidth}
\end{center}
\caption{Sensitivity measures $F$ for each nucleus $(Z,N)$ in the network, for three binding 
energy sensitivity studies using FRDM\cite{mol95} (top panel), DZ\cite{duf95} (middle panel), 
and HFB-21\cite{gor10} (bottom panel) masses.  All three studies use binding energy variations 
of $\pm1$ MeV and astrophysical conditions as used in Ref.~\refcite{bre12}.  Overlaid are 
estimated $10^{-4}$ fission yields for CARIBU\cite{sav05} (thin black line) and FRIB\cite{tar05} 
(wide grey line).}
\label{fig:be}
\end{figure}

One stated aim of the study in Ref.~\refcite{bre12} was to highlight individual nuclear masses 
important for the r-process that are within reach of current and next generation experiments.  
The sensitivity study results presented in Fig.~3 of Ref.~\refcite{bre12} do not give the full 
picture, however, since each separation energy depends on two different nuclear masses.  
Therefore here we repeat this study, varying individual {\em masses} rather than separation 
energies.  Once the mass is varied, the two separation energies that depend on that mass are 
adjusted accordingly, and the r-process baseline simulation is repeated with this variation in 
the nuclear data.  The results are presented in Fig.~\ref{fig:be}.

Figure \ref{fig:be} shows the same bulk features as Fig.~3 of Ref.~\refcite{bre12}.  The largest sensitivity measures 
$F$ are produced by variations in the masses of nuclei along and near the r-process path, particularly at 
the closed shells.  While equilibrium persists between neutron captures and photodissociations, the masses 
determine abundances along each isotopic chain, and the relative abundances of the isotopic chains are set 
by beta decays.  A change to a mass along the chain has the biggest impact on the overall r-process 
abundance pattern if it alters the location of the path and additionally the rate at which material moves 
out of the chain via beta decay. This is described in more detail in Ref.~\refcite{bre12}.

The major difference between Fig.~\ref{fig:be} and Fig.~3 of Ref.~\refcite{bre12} concerns the results for nuclei 
closer to stability than the equilibrium r-process path.  These nuclei are populated at later times in the 
r-process, as $(n,\gamma)$-$(\gamma,n)$ equilibrium fails, material moves toward stability, and neutron 
captures, photodissociations, and beta decays all compete to shape the final abundance pattern.  A 
variation in a neutron separation energy produces a variation in the photodissociation rate 
(Eq.~(\ref{eq:gn})); this can alter the late time nuclear flow, particularly if the nucleus with the varied 
rate is populated and has fallen out of equilibrium.  Ref.~\refcite{sur09} describes this late-time 
photodissociation effect.  Since odd-$N$ nuclei tend to fall out of equilibrium much earlier than even-$N$ 
nuclei, their photodissociation rates, and thus their neutron separation energies, tend to be more 
important at these late times. The sensitivity measures in Fig.~3 of Ref.~\refcite{bre12} showed just this 
expected odd-even behavior.  It is missing from Fig.~\ref{fig:be}, however, because in the binding energy sensitivity 
study a variation in an odd-$N$ separation energy is produced when the binding energy of the odd-$N$ 
nucleus or its adjacent even-$N$ neighbor is altered.  Thus while Fig.~3 of Ref.~\refcite{bre12} highlights 
an interesting feature of r-process dynamics, the new Fig.~\ref{fig:be} more accurately gauges the potential import of 
individual mass measurements on the r-process pattern.

\section{Dependence on astrophysical conditions}\label{sec:astro}

Given the mechanisms by which individual pieces of data influence the r-process abundance pattern described 
in Refs.~\refcite{beu08,sur09,sur11,mum12,bre12,cas12} and briefly above, it is expected that the 
sensitivity measures $F$ will be strongly dependent upon the baseline r-process astrophysical conditions 
chosen for the study.  For example, consider the case outlined in Sec.~\ref{sec:be} above. The equilibrium mechanism 
operates most strongly along the r-process path, which is set by the temperature and neutron number 
density.  Repeating the study with different temperature and density conditions should therefore shift the 
nuclei with the highest sensitivity measures $F$ to correspond to the new path.  The late-time 
photodissociation effect operates as equilibrium is failing but before the temperature has dropped so much 
that photodissociation is shut off entirely.  The range of nuclei for which this effect produces noticeable 
changes to the final abundance pattern is thus sensitively tied to the late-time evolution of the 
temperature and density\cite{mum12b}.

\begin{figure}[ht]
\begin{minipage}[b]{0.48\linewidth}
\psfig{file=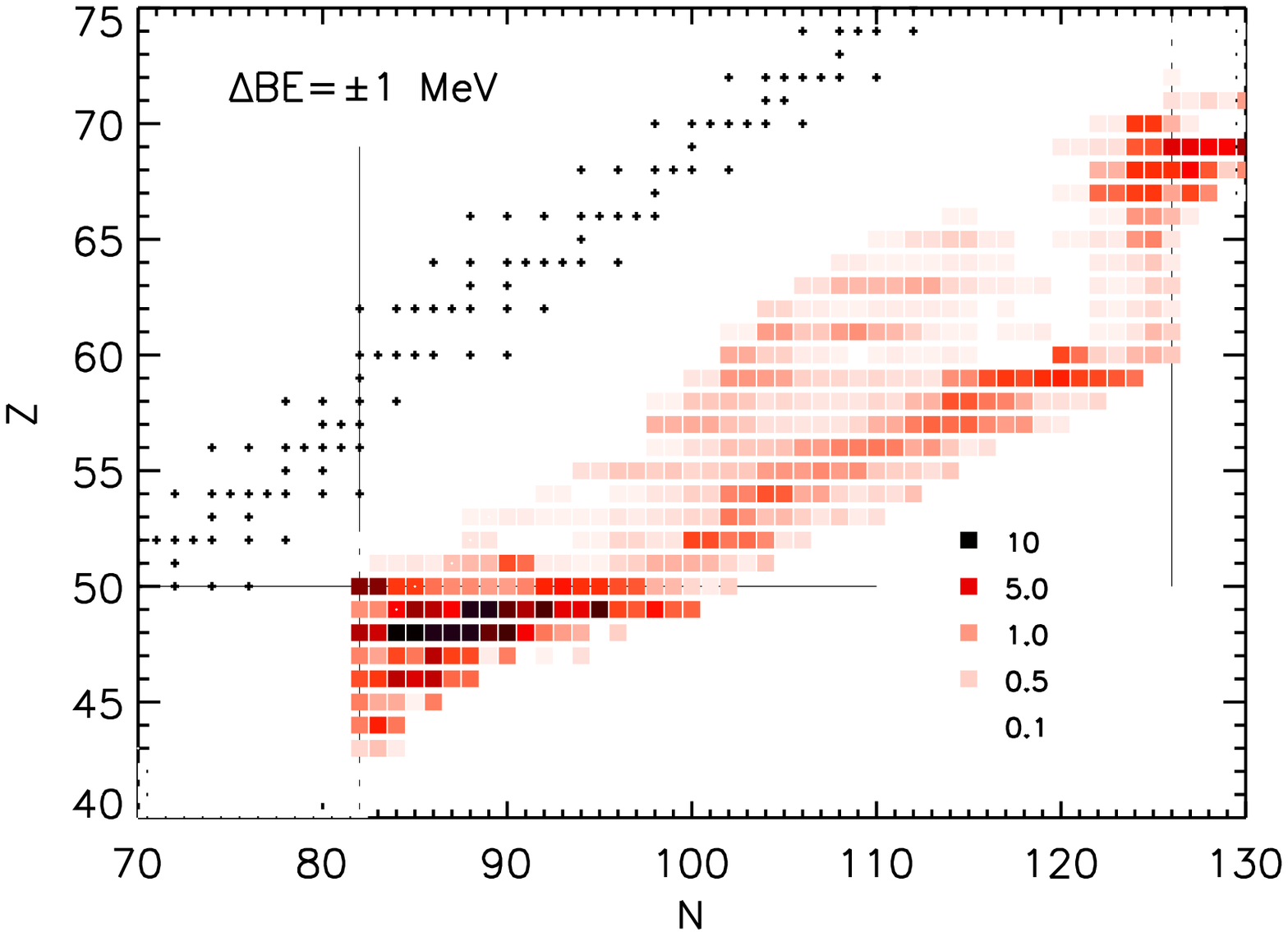,width=\textwidth}
\end{minipage}
\begin{minipage}[b]{0.48\linewidth}
\psfig{file=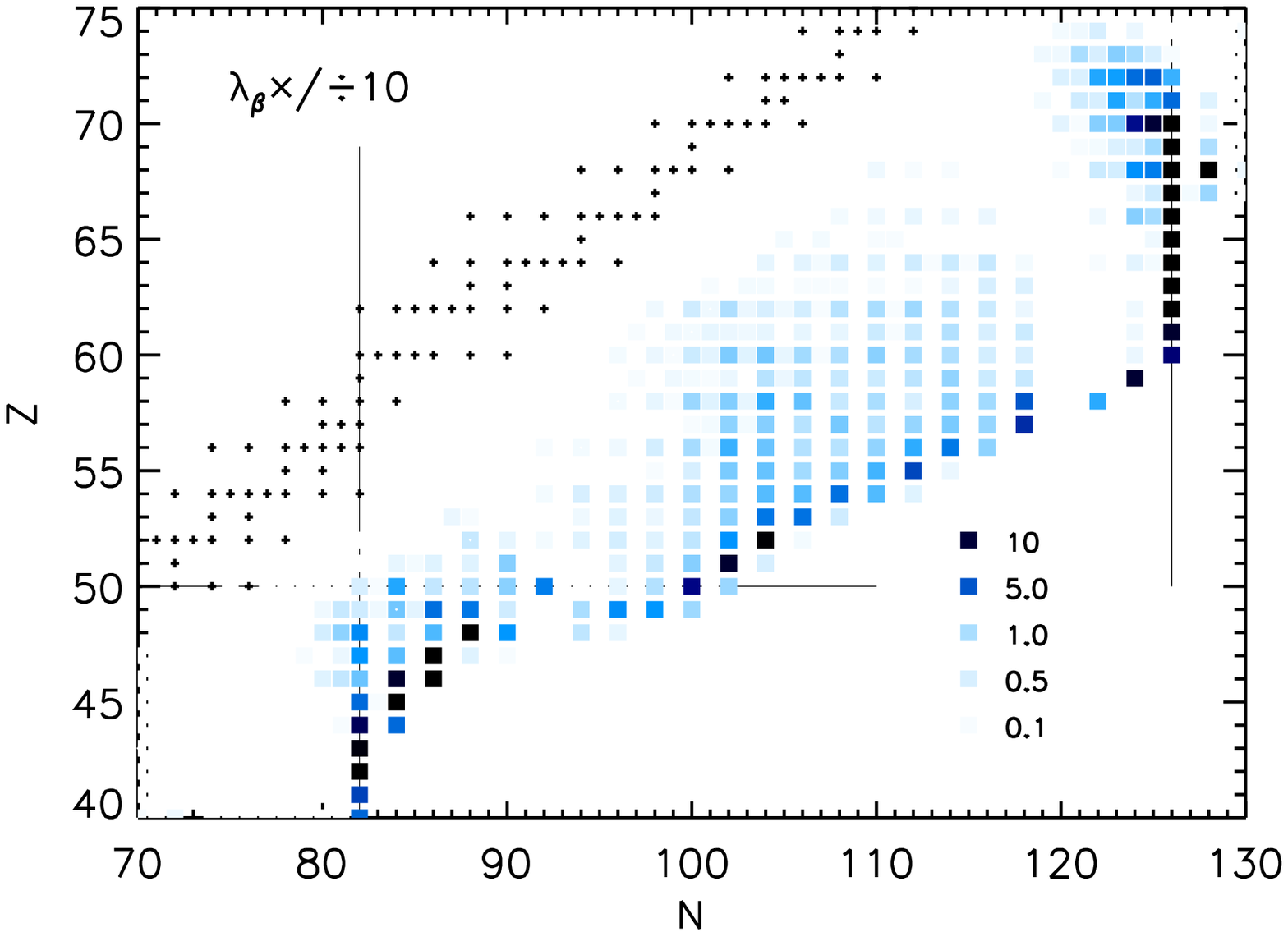,width=\textwidth}
\end{minipage}
\caption{Sensitivity measures $F$ for each nucleus $(Z,N)$ in the network, for the binding 
energy (left) and beta decay (right) sensitivity studies described in Sec.~\ref{sec:astro}.  The studies use 
FRDM masses and a wind parameterization from Ref.~\refcite{mum12} with entropy $s/k=10$ and 
initial electron fraction $Y_{e}=0.150$.}
\vspace{0.3cm}
\begin{minipage}[b]{0.48\linewidth}
\psfig{file=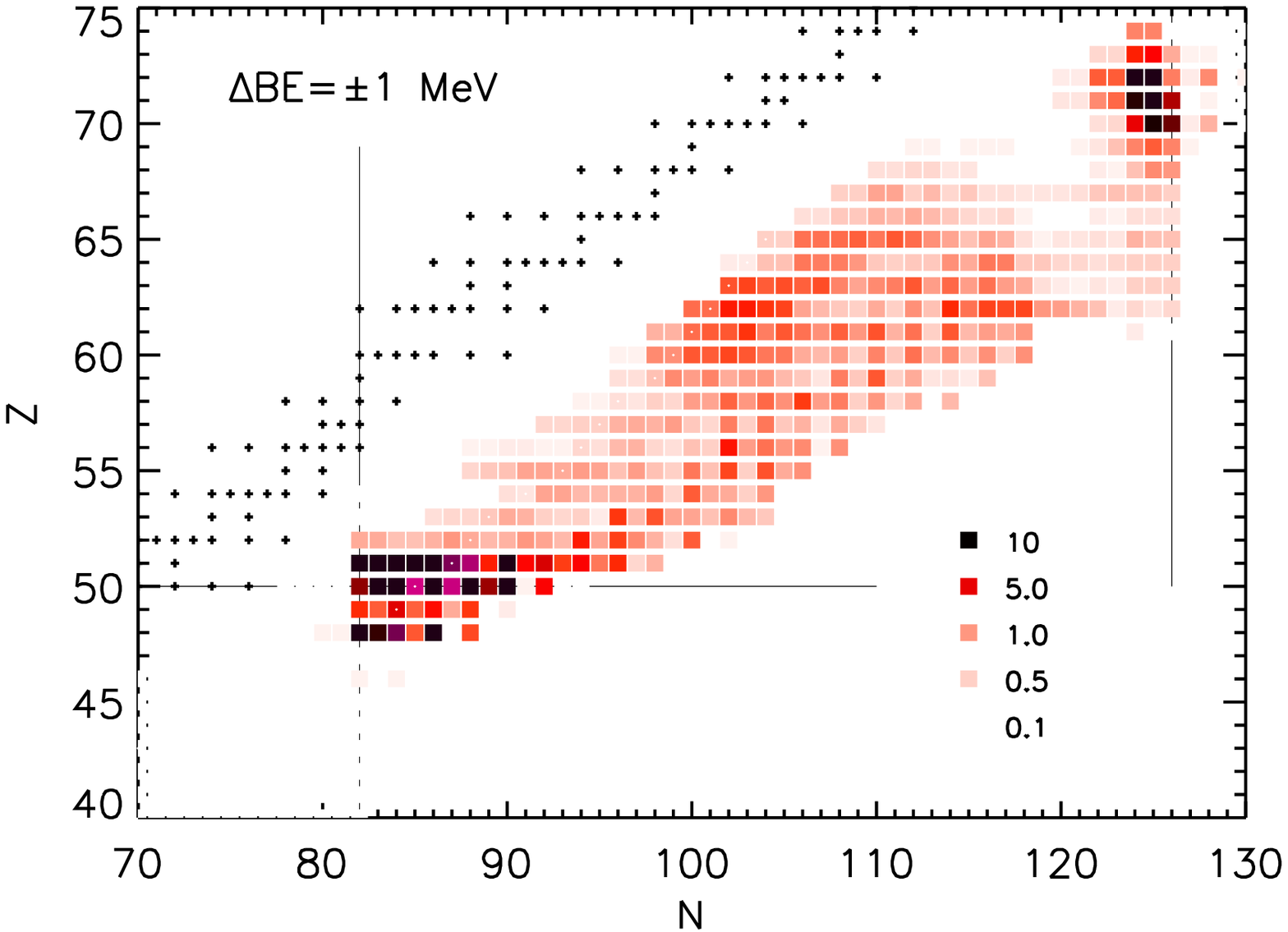,width=\textwidth}
\end{minipage}
\begin{minipage}[b]{0.48\linewidth}
\psfig{file=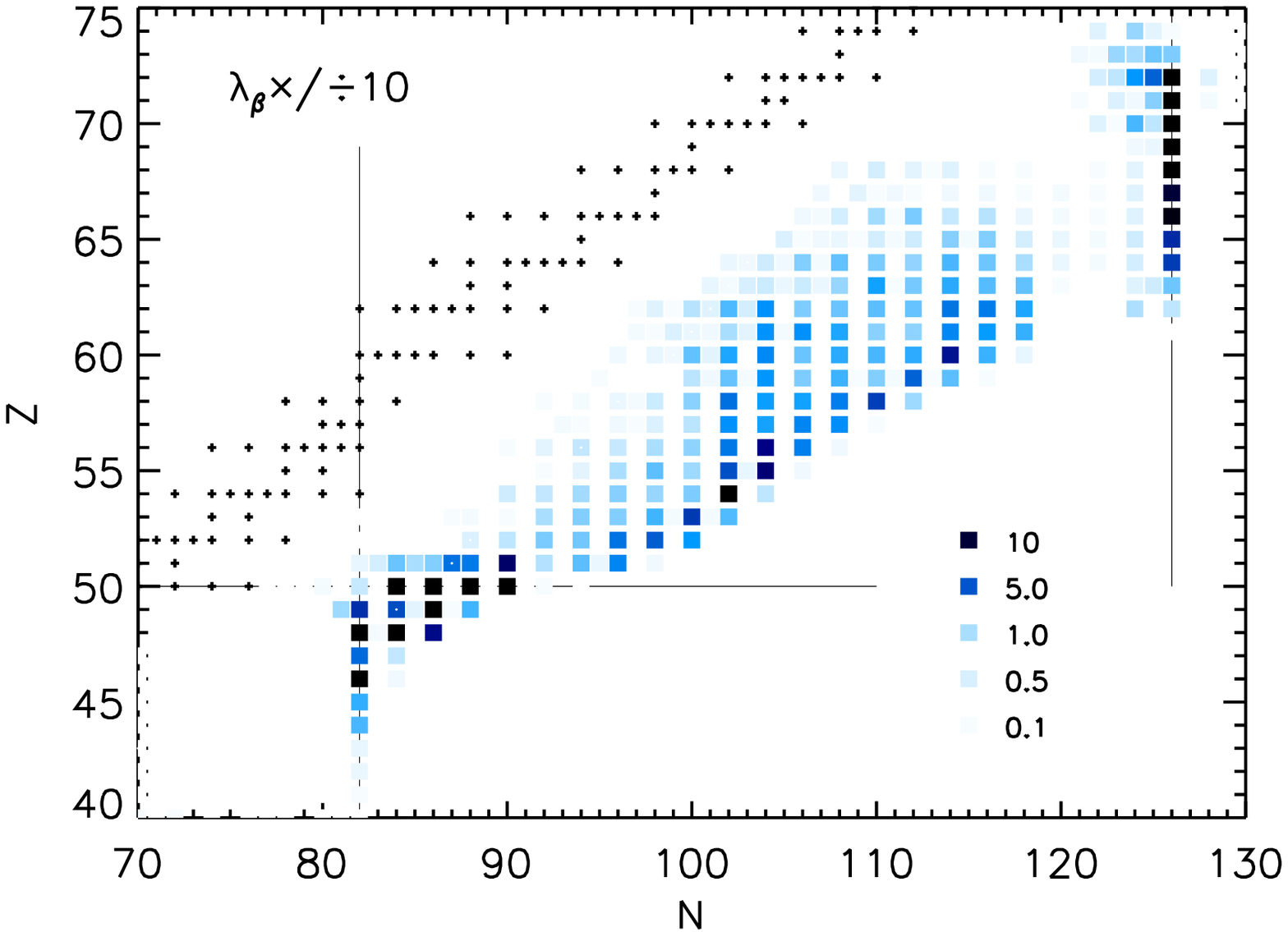,width=\textwidth}
\end{minipage}
\caption{Similar to Fig.~2, except with wind parameters $s/k=100$ and $Y_{e}=0.250$.}
\vspace{0.3cm}
\begin{minipage}[b]{0.48\linewidth}
\psfig{file=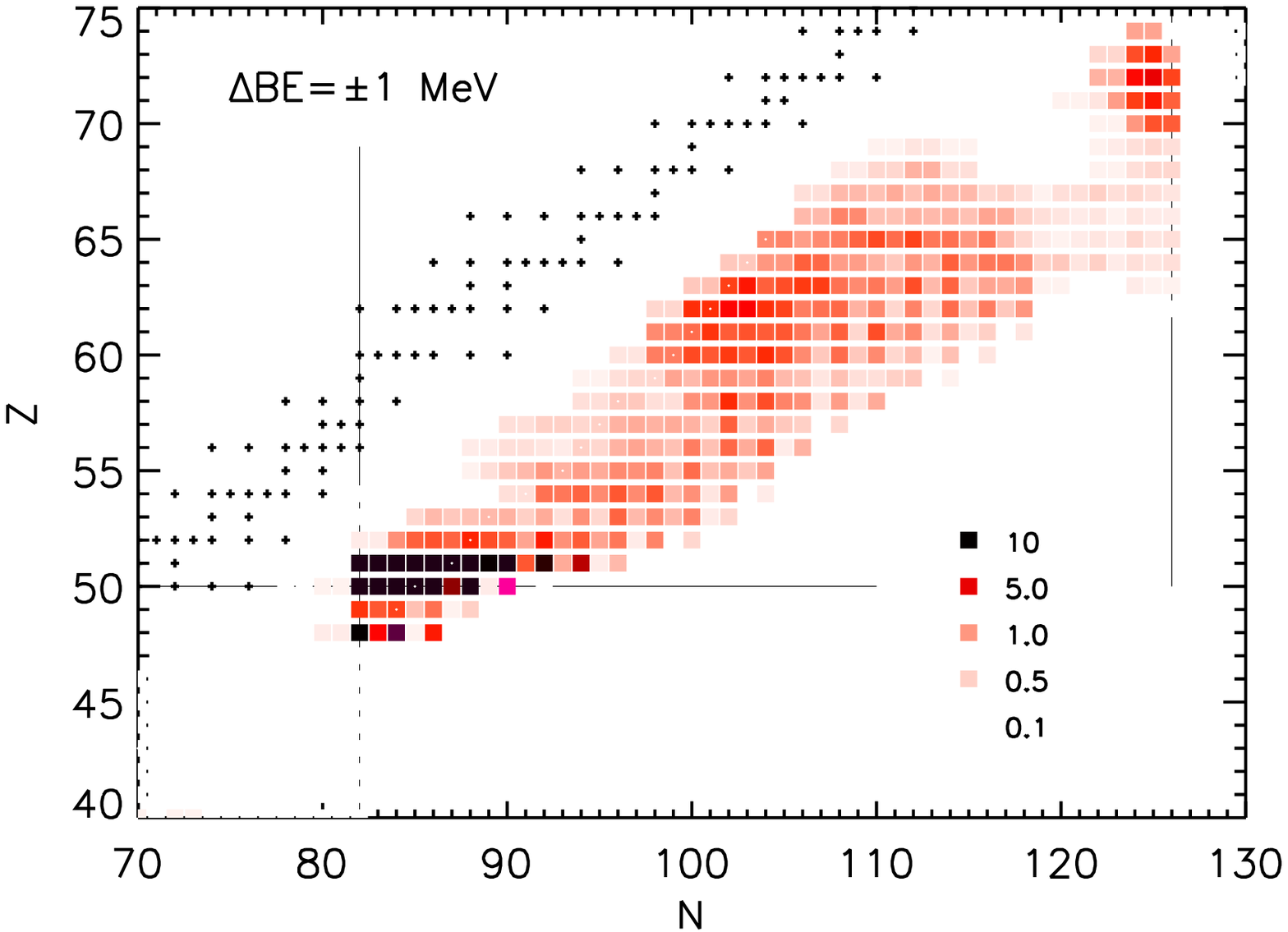,width=\textwidth}
\end{minipage}
\begin{minipage}[b]{0.48\linewidth}
\psfig{file=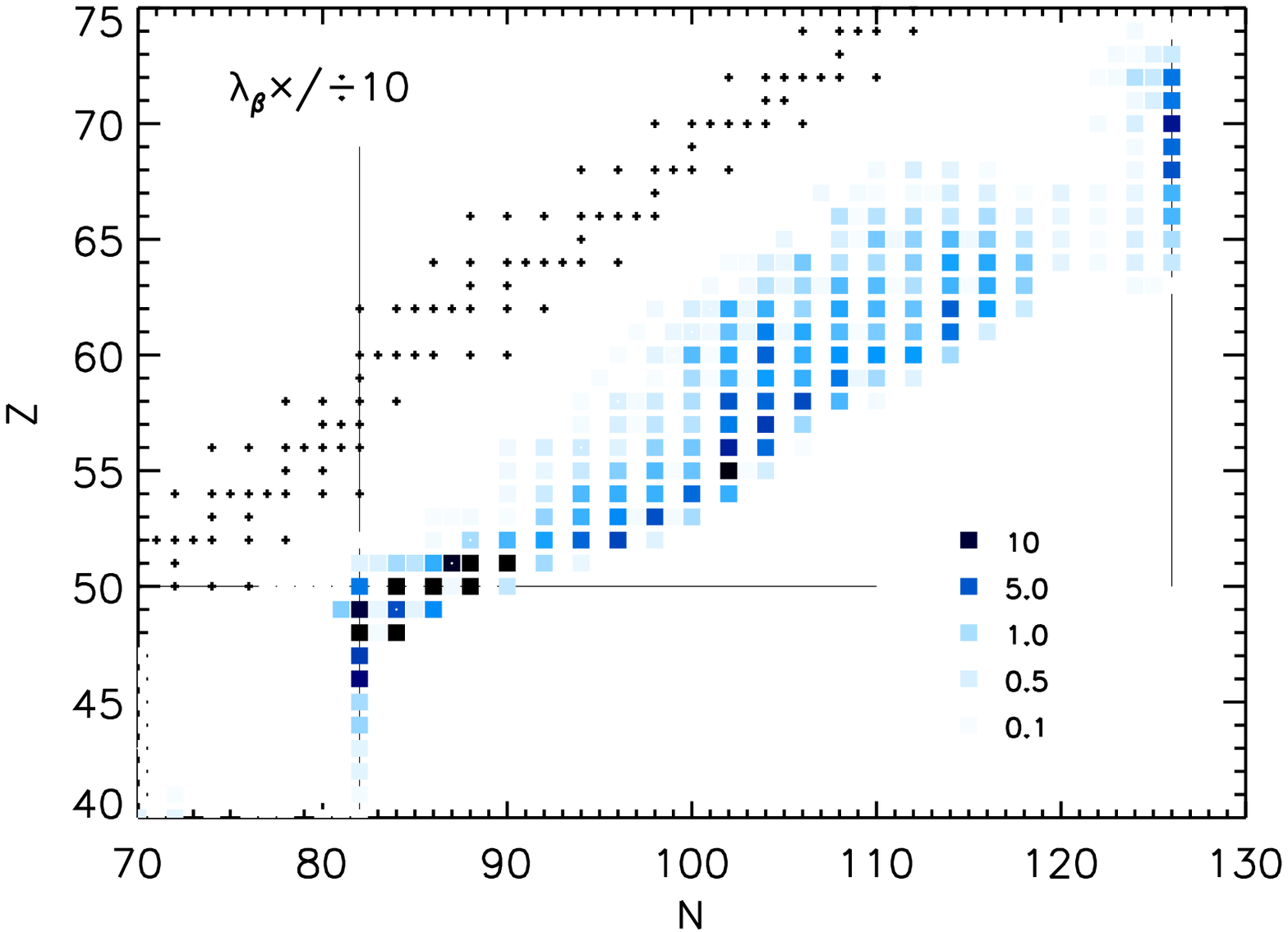,width=\textwidth}
\end{minipage}
\caption{Similar to Fig.~2, except with wind parameters $s/k=200$ and $Y_{e}=0.300$.}
\label{fig:series}
\end{figure}

To examine these effects, we repeat the sensitivity study of Sec.~\ref{sec:be} three times, varying the baseline 
astrophysical conditions each time.  We choose the adiabatic wind parameterization of Ref.~\refcite{mey02} 
as implemented in Ref.~\refcite{mum12b}, where the density as a function of time is given by:
\begin{equation}
\rho(t)=\rho_{1}\exp(-t/\tau)+\rho_{2}\left(\frac{\Delta}{\Delta+t}\right)^{2},
\label{eq:rho}
\end{equation}
where $\rho_{1}+\rho_{2}$ is the density at time $t=0$, $3\tau=\tau_{\mathrm{dyn}}$, and $\Delta$ is a 
constant real number.  Our choices for these parameters here are $\rho(0)=1.58\times 10^{8}$ g/cm$^{3}$, 
$\tau_{\mathrm{dyn}}=80$ ms, and $\Delta=10$ ms.  The temperature is determined from the density and the 
choice of entropy, and the initial composition is determined by the choice of electron fraction $Y_{e}$.  
Here we select three combinations of entropy and initial electron fraction that produce a main r-process 
pattern similar to solar as our baseline simulations for the sensitivity studies.  The results are shown 
in the left panels of Figs.~2-4.

At a given density, a lower entropy trajectory has a lower temperature than a higher entropy trajectory.  
In addition, lower entropy trajectories requires a greater neutron richness to make a successful main 
r-process.  Both of these effects act in the same direction, such that the lowest entropy trajectory 
considered has a path farthest from stability, while the path for the highest entropy trajectory is 
closest to stability.  As a result, the greatest $F$ measures in the binding energy study show the same 
behavior---they are shifted far from stability for the low entropy trajectory (Fig.~2) and closer to 
stability in the high entropy trajectory (Fig.~4).  In addition, higher temperatures persist longer in the 
higher entropy trajectory, so photodissociation rates continue to be important even quite close to 
stability.

In addition to repeating the binding energy study, for the same hydrodynamics we run three beta decay 
sensitivity studies as in Ref.~\refcite{cas12}.  Here the studies proceed as described in Sec.~\ref{sec:sens}, for 
variations in the beta decay rates by a factor of 10.  The results are shown in the right panels of 
Figs.~2-4.  Beta decay sensitivity measures $F$ tend to follow the abundances quite closely\cite{cas12}, 
with the highest $F$ measures generally seen for nuclei along the equilibrium r-process path.  Thus, the 
greatest sensitivity measures evolve with entropy as described above.  However, unlike the 
photodissociation rates that essentially turn off at low temperature, beta decay rates continue to be 
important through even later times, for as long as they compete with neutron capture.  Thus individual beta 
decay rates quite close to stability are influential in all three choices of r-process conditions.

\section{Conclusion}\label{sec:concl}

Sensitivity studies are useful tools to highlight pieces of nuclear data that are important for 
astrophysical processes.  Here we have presented a new binding energy sensitivity study for the r-process 
and examined the impact of variations in astrophysical conditions on the results of this and a beta-decay 
sensitivity study.  We note that a large number of the same nuclei show up as important, particularly at 
the closed shells and in the rare earth region close to stability, despite the differences in 
astrophysical conditions.  Many of these nuclei are within the reach of current radioactive beam 
experiments.

\section*{Acknowledgements}

This work was supported by the National Science Foundation under grant number PHY1068192 and 
through the Joint Institute for Nuclear Astrophysics grant number PHY0822648, and the 
Department of Energy under contract DE-FG02-05ER41398 (RS).

\end{document}